\documentclass[]{aa}  

\usepackage{graphicx}
\usepackage[version=4]{mhchem}
\usepackage{txfonts}
\usepackage{float}
\usepackage{subcaption}
\usepackage{bm}
\usepackage[colorlinks=true,urlcolor=blue,citecolor=blue,linkcolor=blue,breaklinks=true]{hyperref}

\usepackage[normalem]{ulem}

\begin{document}

   \title{Observing Circumplanetary Disks with METIS}

   \author{N. Oberg
          \inst{1,2}
          \and
          I. Kamp 
          \inst{2} 
          \and
          S. Cazaux
          \inst{1,3}
          \and
          Ch. Rab
          \inst{4,5}
          \and
          O. Czoske
          \inst{6}
          }

   \institute{Faculty of Aerospace Engineering, Delft University of Technology, Delft, The   Netherlands  \\
              \email{oberg@astro.rug.nl} 
         \and Kapteyn Astronomical Institute, University of Groningen, P.O. Box 800, 9700 AV Groningen, The Netherlands
         \and
             University of Leiden, P.O. Box 9513, 2300 RA, Leiden, The Netherlands     
         \and
             Universit\"ats-Sternwarte, Fakult\"at f\"ur Physik,   Ludwig-Maximilians-Universit\"at M\"unchen, \\ Scheinerstr.~1, 81679 M\"unchen, Germany\
         \and
             Max-Planck-Institut f\"ur extraterrestrische Physik, Gie{\ss}enbachstra{\ss}e 1, 85748 Garching, Germany 
         \and Institut f\"ur Astrophysik, Universität Wien, Türkenschanzstra{\ss}e 17, 1180 Wien, Austria 
             }

   \date{Received --- accepted ---}

 
  \abstract
   {A full understanding of the planet and moon formation process requires observations that probe the circumplanetary environment of accreting giant planets. The mid-infrared ELT imager and spectrograph (METIS) will provide a unique capability to detect warm-gas emission lines from circumplanetary disks.}
   {We aim to demonstrate the capability of the METIS instrument on the Extremely Large Telescope (ELT) to detect circumplanetary disks (CPDs) with fundamental v=1-0 transitions of $^{12}$CO from 4.5-5 \textmu m.    }
   {We consider the case of the well-studied HD 100546 pre-transitional disk to inform our disk modeling approach. We use the radiation-thermochemical disk modeling code ProDiMo to produce synthetic spectral channel maps.  The observational simulator SimMETIS is employed to produce realistic data products with the integral field spectroscopic (IFU) mode.}
   {The detectability of the CPD depends strongly on the level of external irradiation and the physical extent of the disk, favoring massive \mbox{($\sim10$ M$_{\rm J}$)} planets and spatially extended disks with radii approaching the planetary Hill radius.  The majority of $^{12}$CO line emission originates from the outer disk surface, and thus the CO line profiles are centrally peaked. The planetary luminosity does not contribute significantly to exciting disk gas line emission.  If CPDs are dust-depleted, the $^{12}$CO line emission is enhanced as external radiation can penetrate deeper into the line emitting region.}
   {UV-bright star systems with pre-transitional disks are ideal candidates to search for CO-emitting CPDs with ELT/METIS. METIS will be able to detect a variety of circumplanetary disks via their fundamental $^{12}$CO ro-vibrational line emission in only 60 s of total detector integration time.}


   \maketitle
%

\section{Introduction}

Observations of circumstellar dust have revealed a zoo of intriguing substructures in planet-forming disks \citep{Andrews2020}, including arms (e.g. \citet{Perez2016,Muto2012,Huang2018}), arcs (e.g. \citet{Casassus2013,Marel2013,Isella2013}), rings, gaps, and cavities (e.g. \citet{Calvet2002, Pietu2006,Quanz2013b,ALMAPartnership2015,Andrews2018,Long2018}).  Several of these features are interpreted to be the result of planet-disk interactions \citep{2012ARA&A..50..211K,vanderMarel2016b, Zhang2018, Andrews2020}.  Disks with large inner cavities or gaps are known as (pre-)transition disks \citep{Espaillat2014}, and are promising targets to search for giant planets embedded within \citep{Strom1989,Robinson2011,Zhu2011}.  Massive planets in gaps are expected to continually accrete gas and dust from the surrounding circumstellar disk \citep{Kley1999,Lubow2006,Morbidelli2014,Teague2019} into a moon-forming  circumplanetary disk (CPD) \citep{Canup2002,2003Icar..163..198M,Szulagyi2016}. 

As of July 2022, direct detection of extrasolar circumplanetary dust associated with an accreting planet has only been confirmed in the PDS 70 system. The PDS 70 circumstellar disk features a gap with a radial width of 70 au \citep{Hashimoto2012} containing two planets.  The first planet PDS 70 b was detected in the NIR \citep{Keppler2018,Muller2018} and in H-$\alpha$ \citep{Wagner2018,Haffert2019}. A second planet PDS 70 c was discovered also in H-$\alpha$ \citep{Haffert2019}.  Unresolved sub-mm dust thermal emission is co-located with this planet, interpreted as originating from a dusty CPD \citep{Isella2019,Benisty2021}.  Self-consistent radiative transfer modeling of the system suggests the CPD is optically thick with an upper limit on the dust mass of 0.7 M$_{\oplus}$ \citep{Bayron2022}, but it is noted that high spatial and spectral resolution observations of the gas component are needed to break degeneracies between the planet and disk properties.

No other (pre-)transitional disk gap has been found to contain circumplanetary dust. \citet{Francis2020} suggest that PDS 70 is anomalous in that the gap may only recently have been opened, with correspondingly high rates of accretion onto the planets within the gap.  Alternatively, episodic accretion may limit the visibility of planets to brief periods, with the relatively inviscid CPD acting as an accretion ``bottleneck" \citep{Lubow2012,Brittain2020}. The rapid depletion of dust due to fast inwards aerodynamic drift in CPDs may also hinder attempts to detect them in continuum emission \citep{Zhu2018,Rab2019}. 

An additional candidate CPD has been detected in the \mbox{AS 209} disk by $^{13}$CO $J$=2-1 gas line emission \citep{Bae2022}.  The CPD candidate candidate is embedded within an annular gap seen in the $^{12}$CO emission at a radial distance of 200 au.  The gap region is still optically thick in $^{12}$CO and hence there is no corresponding detection in $^{12}$CO.  However, a perturbation in the velocity field of the $^{12}$CO gas and evidence of localized heating further supports the interpretation of a planet+CPD.  Assuming a standard $^{13}$CO abundance, the gas mass of the CPD is estimated to be 30 M$_{\oplus}$.  Given the nondetection of associated continuum emission the dust mass of the CPD must be \mbox{$<0.027$ M$_{\oplus}$}, suggesting a dust-to-gas ratio \mbox{$<9\times10^{-4}$}, in line with prediction of rapid dust depletion of wide-separation CPDs \citep{Zhu2018, Rab2019}.  Gas line observations thus provide a way to detect CPDs even if they are strongly dust depleted.

\subsection{HD 100546}

One of the best studied pre-transitional disks surrounds the Herbig star HD\,100546 \citep{Ancker1997,Vioque2018}.  The system lies at a distance \mbox{108.1 $\pm$ 0.5 pc}, right ascension  11h 33m 25.3s, declination -70$^{\circ}$ 11' 41.2$^{\prime\prime}$ \citep{Brown2021}. The age is estimated to be \mbox{7.02 $\pm$ 1.49 Myr}  \citep{Fairlamb2015} or 5.5$^{+1.4}_{-0.8}$ Myr \citep{Vioque2018}. The system is comprised of a clearly divided inner and outer disk, with a gap depleted in gas and dust radially spanning \mbox{$\sim$1-20 au} \citep{Bouwman2003,Grady2005,Brittain2009,Avenhaus2014,Fedele2015,Jamialahmadi_2018,Pineda2019}.  

The HD 100546 system hosts several potential planet candidates, either claimed via direct detection or indirectly via their influence on the circumstellar disk structure and dynamics. The first candidate companion "b" was identified by direct imaging at \mbox{3.8 \textmu m} at a separation \mbox{0.48$\arcsec \pm$0.04} (projected separation $\sim$70 au) \citep{Quanz2013,Currie2014,Quanz2015}.  However the existence of this object has been called into question as potentially being an artifact of data reduction \citep{Rameau2017,Cugno2019}. 

A second closer-in companion "c" has been tentatively detected inside the cavity during multi-epoch monitoring of spatially unresolved CO (v=1-0) line profile asymmetries potentially originating from the gas component of a CPD inside the cavity \citep{Brittain2013II,Brittain2014}. This claim has also been disputed, with the line asymmetry being attributed to e.g. a slit misalignment \citep{Fedele2015}. However \citet{Brittain2019} rebuts this claim, arguing that a slit misalignment is not plausible given that the CO (v=1-0) and hot band lines were observed simultaneously yet displayed differing spectro-astrometric signals and profiles from 2006 to 2013. Furthermore the signal attributed to the CPD appears to vanish when the predicted position of the planet moves behind the cavity inner edge \citep{Brittain2019}. Evidence for the ``c" companion within the gap at 10-15 au has been further strengthened by modeling efforts that reproduce the observed mm-dust disk substructure with planet-disk interaction \citep{Pinilla2015,Ackermann2021,Fedele2021}.  In this work we focus on the direct detection of the CPD of this potential planet candidate "c" by $^{12}$CO (v=1-0) line emission.

\subsection{CO line observations with ELT/METIS}

The Extremely Large Telescope (ELT) is a next generation observatory under construction at the peak of Cerro Armazones in the Atacama desert of Chile.  With a segmented primary mirror diameter of 39.3 meters the sensitivity and angular resolution of the ELT promises to significantly bolster the capabilities of ground-based IR astronomy \citep{Ramsay2018}.  The Mid-infrared ELT Imager and Spectrograph (METIS) is a planned instrument designed for the observation of exoplanets and protoplanetary disks \citep{Brandl2021}.  The single-conjugate adaptive optics system allows METIS to perform high-contrast diffraction-limited integral field unit (IFU) spectroscopy at a spectral resolution $R\sim10^5$ in the L and M bands. This offers the capability to observe the fundamental transitions of \mbox{$^{12}$CO (v=1-0)} from \mbox{4.5-5 \textmu m}.  This CO emission primarily traces warm gas such as at the inner rim of the disk gap above the midplane, and has been used to deduce the presence of cavities or gaps \citep{Brittain2009,Banzatti2015,Hein2016,Antonellini2020}. A CPD inside a disk gap may be exposed to significant scattered stellar radiation \citep{Turner2012, Oberg2020}, heating gas in the CO line emitting region of the CPD and increasing  emission.

Analysis of the emission allows for the kinematic derivation of a planet's mass \citep{Rab2019}.  Limits on the CPD gas temperature, composition, physical extent, and total mass may also be determined.  High spectral resolution observations in the Mid- and NIR can thus provide clues to determine the properties of the planet, the process of gas giant accretion, and the  formation of regular satellite systems around massive planets.  If CPDs are strongly dust-depleted and accretion onto the planet is only episodic, gas line observations may offer the most promising avenue to probe the circumplanetary environment.

We performed simulations to demonstrate the METIS instrument capabilities will enable the detection of a planet+CPD located in the gap of (pre-)transitional disks, and that high spectral resolution IFU spectroscopy of CO ro-vibrational emission will enable an unprecedented insight into the nature of these objects. The outline of this work is as follows: in Sect. \ref{sec:methods} we describe the capabilities of the disk modeling code,  the properties of the disk model,  and capabilities of the telescope observing simulation tools.  In Sect. \ref{sec:results} we describe our results, and discuss implications and conclusions in sections \ref{sec:discussion} and \ref{sec:conclusions}, respectively.

\begin{figure}
  \includegraphics[width=\textwidth/2]{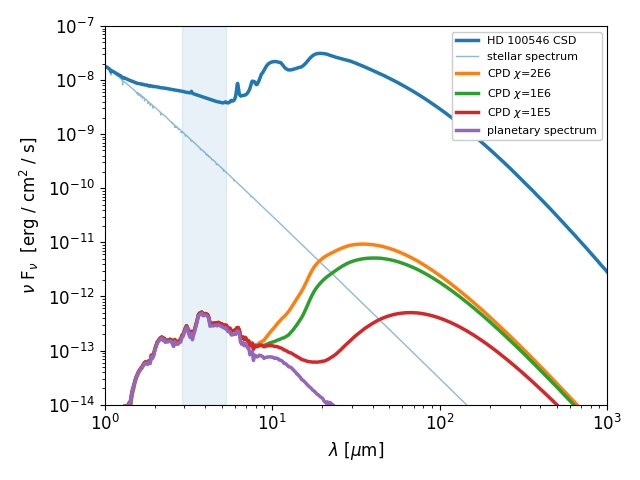}
      \caption{Continuum SED of the HD 100546 circumstellar disk (blue line) and of the CPD for three cases of differing background radiation (orange, green, and red lines) and in the case of no CPD (purple line) for a planet with $T_{\rm eff} = 1000$ K.  The HD 100546 stellar spectrum is also included (thin blue line). The METIS LMS wavelength range is indicated by the light blue filled region.}
      \label{fig:sed-continuum}
\end{figure}

\begin{figure}
  \includegraphics[width=\textwidth/2]{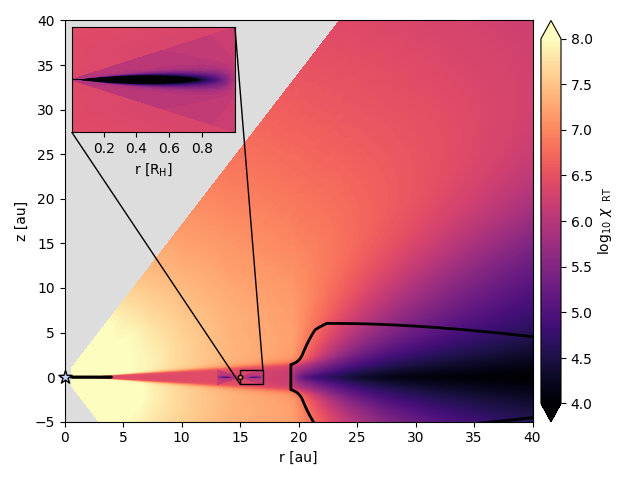}
      \caption{Geometry and intensity of the UV field strength $\chi_{\rm RT}$ at each position in the CSD and CPD derived from the results of the 2D radiative transfer.  The CPD model UV field map is overlaid for illustrative purposes.  The black contours trace the surface of minimum optical extinction A$_{\rm V}$=1 in the vertical or radial direction.}
      \label{fig:chi-shadow}
\end{figure}

\section{Methods} \label{sec:methods}

To model the disk physics and chemistry and to produce synthetic line emission data cubes of the HD 100546 circumstellar disk (CSD) and candidate CPD we used the radiation thermochemical disk modeling code \textsc{ProDiMo} (Protoplanetary Disk Model) \footnote{\url{https://prodimo.iwf.oeaw.ac.at/}}  \citep{Woitke2016,Kamp2017,Woitke2019,Thi2020H2}. The CO abundance in the disks is self-consistently calculated with a rate-based approach using the 'large' chemical network including 235 species and 13 elements described in \citet{Kamp2017}.  The adopted elemental abundances of C and O are 1.38$\times 10^{-4}$ and 3.02$\times10^{-4}$ relative to H, respectively \citep{Savage1996}. \textsc{ProDiMo} has been used previously to model observations of CO line emission from CPDs around wide-orbit companions in the sub-mm \citep{Rab2019}.  To model CO emission from Herbig disks \textsc{ProDiMo} utilizes a CO molecule model that includes up to 50 rotational levels and 9 vibrational levels for the electronic ground state \mbox{X$^1\Sigma^+$} and the first excited electronic state \mbox{A$^1\Pi$}. Collisions between CO molecules and hydrogen molecules (H$_2$), hydrogen atoms (H), helium (He), and electrons are taken into account \citep{Thi2013,Song2015}.  Order-of-magnitude uncertainties in rate coefficients result in model CO line flux variations of up to 20$\%$.

\subsection{System model properties}

In this section we describe the properties of the circumstellar and circumplanetary disk models. We assumed that the CSD and CPD are coplanar ($i_{\rm CSD}$ = $i_{\rm CPD}$), that the CPD lies within the midplane of the CSD on a zero-inclination orbit, and that the CPD and CSD are co-rotating (rotating in the same direction).  The CPD is assumed to lie on a circular orbit ($e_{\rm CPD} = 0$) at a distance of 15 au from the star. For the production of combined CPD+CSD data cubes the position angle of the CPD is varied from 0-180$^{\circ}$ in steps of 45$^{\circ}$ relative to the total system position angle of $\sim 140^{\circ}$.   The continuum SED of the CSD and CPD for several of the models can be found in Fig. \ref{fig:sed-continuum}.

\subsubsection{Properties of the HD 100546 Circumstellar Disk} \label{sec:HD100546-model}

The properties of the HD 100546 disk have previously been derived by spectral energy distribution (SED) fitting by means of a genetic algorithm \citep{Woitke2019}.  A multi-wavelength set of publicly available photometric fluxes, low- and high resolution spectra, and interferometric data have been collated to produce a global SED \citep{Dionatos2019}.  Sixteen free parameters describing the disk physical and chemical parameters have been fit to the SED by iteratively performing MCFOST radiative transfer simulations.  Details of the disk modeling procedure, SED fitting, and limitations of the SED fitting process can be found in \citet{Woitke2016}, \citet{Kamp2017},\citet{Woitke2019} and \citet{Dionatos2019}.   The disk gap is parameterized to span \mbox{4-19.3 au}.  A detailed description of the model parameters can be found in \citet{Roche2021} \footnote{The model parameters and output can also be accessed directly at \url{http://www-star.st-and.ac.uk/~pw31/DIANA/SEDfit/HD100546_model_index.html}}.  Inside the gap the vertical gas column density does not exceed $N_{\langle \rm H  \rangle} = 10^{17}$ cm$^{-2}$. However in the following section we considered also the implications of additional gas and dust being present in the gap.

\begin{figure*}
  \includegraphics[width=\textwidth]{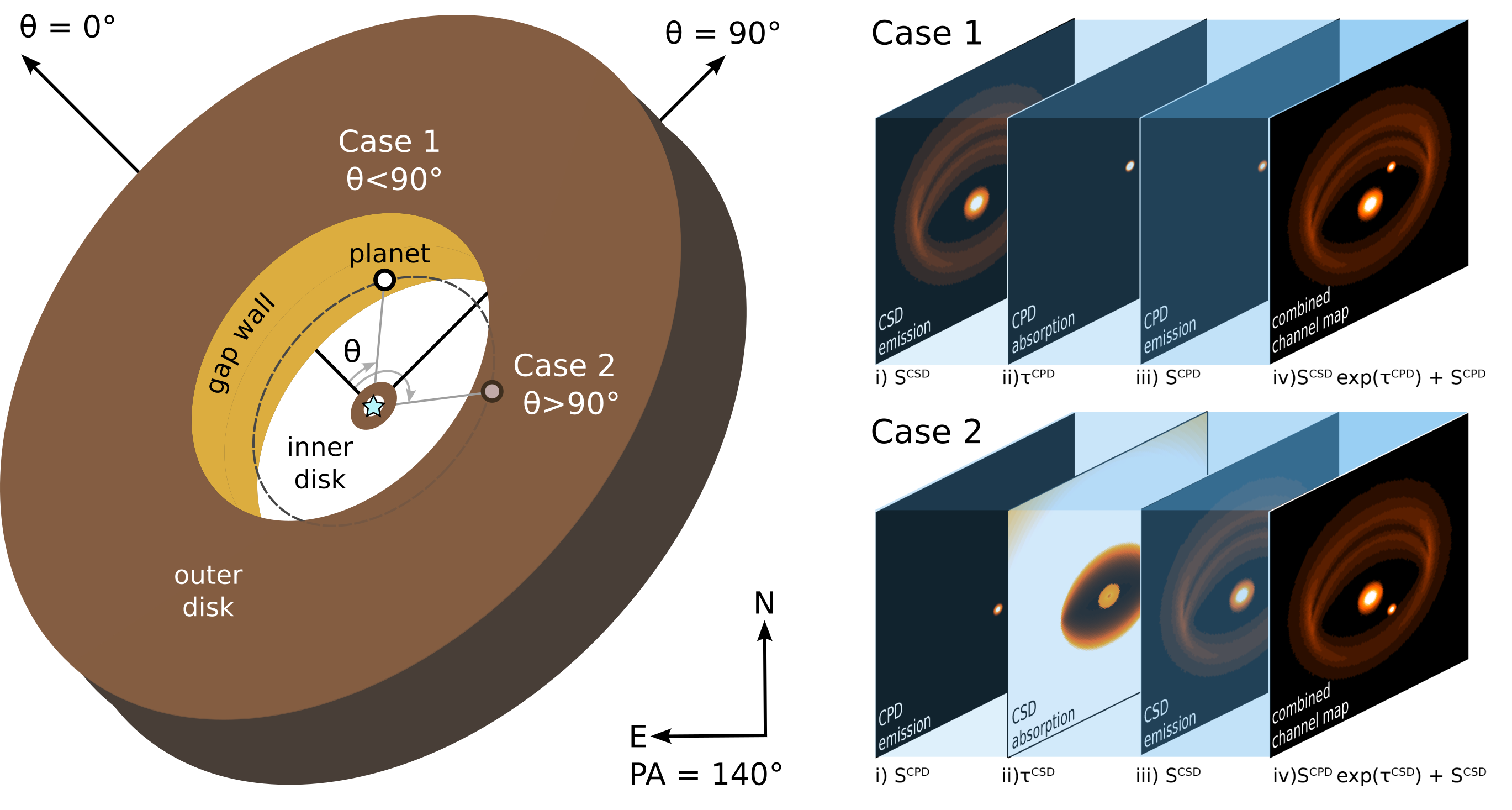}
      \caption{Illustration of the circumstellar disk geometry and the components contributing to the channel maps. In case 1 the CPD phase angle \mbox{$\theta$ is < 90$^{\circ}$} (CPD as foreground object relative to the CSD).  The CPD absorption map (ii) is used to extinct the CSD emission map (i).  The CPD emission (iii) is then added to produce the final image (iv). When \mbox{$\theta$ > 90°} the CSD gap edge becomes the foreground object, and the process follows case 2. At \mbox{$\theta$ = 90$^{\circ}$} the extinction caused by either object on the other is negligible and the two channel maps are simply co-added.}
      \label{fig:cube-combiner}
\end{figure*}

\subsubsection{Properties of the planet and CPD} \label{sec:cpd-chi}

We considered two cases for the physical size of the CPD. The first reference case is one in which the outer radius $R_{\rm out}$ of the CPD is equal to the planetary Hill radius R$_{\rm H}$.  In the second case we assume that the CPD is truncated to one third of the Hill radius due to tidal interaction or photoevaporative effects \mbox{($R_{\rm out}$ = 1/3 R$_{\rm H}$)} \citep{2011MNRAS.413.1447M,2011AJ....142..168M, Oberg2020}.  For the reference model we assume a planetary mass of 10 M$_{\rm J}$, but consider also the case of a 5, 3 and 1 M$_{\rm J}$ planet with correspondingly smaller Hill radii.  An exponential decline in the surface density profile of the CPD is parameterized to begin at one third of the outer radius in either case.  The planetary luminosity is set to be $10^{-5}$ L$_{\odot}$ to be consistent with the predicted post-runaway accretion phase of giant planet evolution \citep{Marley2007}. The planetary spectrum is adopted from the DRIFT-PHOENIX library for an object with $T_{\rm eff} = 1000$ K \citep{Helling2008}. 

Given the short timescales of dust depletion in the CPD \citep{Zhu2018,Rab2019} and the possibility that pressure bumps at the gap edge prevent significant transport of dust towards the CPD \citep{Rice2006,Zhu2012} we adopt a relatively dust depleted CPD with a dust-to-gas ratio $d/g = 10^{-3}$, but find that adopting a canonical value of 10$^{-2}$ has a negligible impact on the ro-vibrational CO line emission characteristics of the CPD.  We consider also more extreme cases of dust depletion with \mbox{$d/g = 10^{-4}$}, $10^{-5}$ and $10^{-6}$.  A summary of the various CPD variations of parameters can be found in Table \ref{tab:cpd-properties}. The parameters which remain fixed and are shared between all models are found in Table \ref{tab:cpds}.

The CPD is externally irradiated by the star located at the center of the circumstellar disk.  The strength of the attenuated UV field $\chi_{\rm RT}$ was calculated throughout the circumstellar disk as part of the full 2D radiative transfer. The value of $\chi_{\rm RT}$ represents the strength of the local UV radiation field with respect to the standard interstellar radiation field \citep{Roellig2007}.  We extracted the intensity of this attenuated stellar radiation at the position of the CPD.  This extracted radiation field was then applied as a background to the CPD.  In practice the external radiation field is represented in the radiative transfer by a diluted 20000 K blackbody component which is isotropically incident on the CPD.  

The geometry of the UV radiation field intensity derived from the 2D radiative transfer in the CSD is depicted in Fig. \ref{fig:chi-shadow}. For illustrative purposes the corresponding radiation field in and around the reference CPD has been scaled and positioned at the stellocentric radius corresponding to its physical separation and extent.  The shadowing effect of the inner disk is apparent, and the CPD sits entirely within this shadow.  In the midplane the $\chi_{\rm RT}$ field strength at \mbox{15 au} is \mbox{1.8$\times10^6$}. When applied as a background UV field to the CPD model the irradiation is assumed to be isotropically incident.  Given the geometry and alignment of the CPD relative to the star and corresponding potential for self-shadowing we test also significantly reduced external UV field strengths of \mbox{$\chi = 10^4$}, $10^5,$ and $10^6$.  We do not consider the possibility of a warped or inclined inner disk, which might periodically vary the magnitude of the inner-disk's shadowing.

We assessed the contribution of back-scattered light by performing a separate radiative transfer simulation where the outer disk component of the HD 100546 CSD model is removed to prevent back-scattering from the gap wall at 19.3 au.  If the outer disk is removed entirely the $\chi_{\rm RT}$ value at the position of the CPD reduces by a factor 20 to $10^5$, hence 95$\%$ of the incident emission is scattered radiation originating from the outer gap wall.  The question of how much external radiation is effectively incident on the CPD is discussed further in Sect. \ref{sec:appendix:gapUV}.

The distribution of gaseous CO in the reference CPD is shown in Fig.\ref{fig:cpd-line-origin}. The 2D gas temperature and density structure of the CPD are shown in Fig. \ref{fig:cpd-tgas} and Fig. \ref{fig:cpd-rhog}, respectively. For the calculation of the line Doppler shift we adopt for the stellar rest frame a radial velocity of 9.25 km s$^{-1}$ \citep{Brown2021}.  The planetary radial velocity at maximum elongation is \mbox{8.15 km s$^{-1}$} for \mbox{$a = 15$ au}, \mbox{$i = 42^{\circ}$}.

\subsection{Radiative transfer post-processing}

The \textsc{ProDiMo} line radiative transfer implementation precludes a full 3D treatment of the combined CPD and CSD system.  Instead, two independent disk models are combined in a post-processing step.  First a 2D axisymmetric radiative transfer is performed independently in the CSD model (see section \ref{sec:HD100546-model}).  We extract the properties of the radiation field in the gap midplane of the CSD model.  These properties were used to inform the plausible magnitude of external UV irradiation of the CPD.  The CPD disk model was then initialized with these background conditions.

\begin{table}
\caption{CPD model variations.  Variations on the reference model are highlighted in bold.}         
\label{tab:cpd-properties}     
\centering                         

\resizebox{0.49\textwidth}{!}{
\begin{tabular}{l c c c c c}       
\hline\hline               
model id & $\chi_{\rm gap}$ & $R_{\rm out}$ [au] & $M_{\rm p}$ [M$_{\rm J}$]  & $M_{\rm CPD}$ [M$_{\rm P}$]  & $d/g$  \\   
\hline   
\vspace{1ex}
   reference     & 2$\times10^6$ & 1.62   & 10 &  10$^{-2}$   & 10$^{-3}$  \\   
   
   chi1E6 & $\bm{1\times10^6}$ & 1.62   & 10 &   10$^{-2}$   & 10$^{-3}$     \\     
   chi5E5 & $\bm{5\times10^5}$ & 1.62   & 10 &   10$^{-2}$   & 10$^{-3}$     \\  
   \vspace{1ex}
   chi1E5 & $\bm{1\times10^5}$ & 1.62   & 10 &   10$^{-2}$   & 10$^{-3}$     \\  
   
   chi2E6-s & $\bm{2\times10^6}$ & \textbf{0.54} & 10 &  10$^{-2}$    &   10$^{-3}$    \\
   chi1E6-s & $\bm{1\times10^6}$ & \textbf{0.54} & 10 &  10$^{-2}$    &   10$^{-3}$    \\ 
   chi5E5-s & $\bm{5\times10^5}$ & \textbf{0.54} & 10 &  10$^{-2}$    &   10$^{-3}$    \\   
   \vspace{1ex}
   chi1E5-s & $\bm{1\times10^5}$ & \textbf{0.54} & 10 &  10$^{-2}$    &   10$^{-3}$    \\     
   
   chi2E6-5mj & 2$\times10^6$ &  \textbf{1.29} & \textbf{5}   &  10$^{-2}$   & 10$^{-3}$  \\      
   chi2E6-3mj & 2$\times10^6$ &  \textbf{1.08} & \textbf{3}   &  10$^{-2}$   & 10$^{-3}$  \\     
   \vspace{1ex}
   chi2E6-1mj & 2$\times10^6$ &  \textbf{0.75} & \textbf{1}   &  10$^{-2}$   & 10$^{-3}$  \\

   mcpd3 & 2$\times10^6$ & 1.62 & 10 &  \textbf{10$^{-3}$}  &  10$^{-3}$  \\
   mcpd4 & 2$\times10^6$ & 1.62 & 10 &  \textbf{10$^{-4}$}  &  10$^{-3}$  \\ 
   \vspace{1ex}
   mcpd5 & 2$\times10^6$ & 1.62 & 10 &  \textbf{10$^{-5}$}  &  10$^{-3}$  \\

   chi2E6-dg4 & 2$\times10^6$ & 1.62   & 10 &   10$^{-2}$  & \textbf{10$^{-4}$}  \\     
   chi2E6-dg5 & 2$\times10^6$ & 1.62   & 10 &   10$^{-2}$  & \textbf{10$^{-5}$}  \\   
   \vspace{1ex}
   chi2E6-dg6 & 2$\times10^6$ & 1.62   & 10 &   10$^{-2}$  & \textbf{10$^{-6}$}  \\     

\hline                                                 
\end{tabular}}
\end{table}

\begin{table}
    \caption{Common parameters shared between all CPD models listed in Table \ref{tab:cpd-properties}.}

    \centering
    \renewcommand{\arraystretch}{1.1}%

   \begin{tabular}{llll}
    \hline \hline
        Parameter               & Symbol              & Value         & Unit          \\ \hline  

        Planetary Luminosity    & $L_{\rm p}$         & $10^{-4}$     &  L$_{\odot}$  \\
        
        Effective Temperature   & $T_{\rm eff,p}$     & 1000          & K           \\
        UV Luminosity           & $L_{\rm UV,p}$      & 0.01          & L$_{\rm p}$ \\
      
        \hline

        Disk Inner Radius         & $R_{\rm in} $    & 0.01   & au \\
        Column Density Power Ind. & $\epsilon$ & 1.0    & -  \\
        Flaring Index             & $\beta$    & 1.15   & -  \\
        Reference Scale Height    & $H_{\rm 0.1 au}$     & 0.01   & au \\
        
        \hline
        Minimum dust size     & $a_{\rm max}$  & 0.05        & \textmu m   \\
        Maximum dust size     & $a_{\rm max}$  & 3000        & \textmu m   \\
        Dust size power law index & $a_{\rm pow}$      & 3.5         & -         \\
       
        Dust composition: \\
         \hspace{0.5cm} Mg$_{0.7}$Fe$_{0.3}$SiO$_3$ &    & $60\%$ \\
         \hspace{0.5cm} Amorphous carbon            &    & $15\%$ \\
         \hspace{0.5cm} Vacuum                      &    & $25\%$ \\
        
    \end{tabular}

    \label{tab:cpds}
    
\end{table}

Line radiative transfer was performed independently for the CSD and CPD, producing  data cubes for selected CO ro-vibrational lines in the METIS LMS-mode wavelength range. In both cases a map of the line-of-sight optical depth ($\tau$ map) through the disks was also produced at each wavelength.  The line data cubes of the CSD and CPD were then combined.  The details of the combination process are dependent on the phase angle $\theta$ of the CPD along its orbit (where $\theta = 0$° corresponds to the northern minor axis of the disk).  For a given $\theta$ the line-of-sight velocity of the CPD is calculated to determine the wavelength shift $\delta \lambda$ of the line center.  At values of $\theta <$ 90° (where 90° corresponds to maximum elongation) the CPD  partially occludes CSD emission originating from the near-side cavity wall. The $\tau$ map was used to calculate the extinction of the background CSD emission due to the CPD dust and gas.  At each pixel $i,j$ in the CSD cube the flux $S$ was extincted at the wavelength $\lambda$ according to 

\begin{equation}
    S_{i,j}(\lambda) = S^{\rm CSD}_{i,j}(\lambda) \, e^{-\tau^{CPD}_{i,j}(\lambda+\delta \lambda)} + S^{\rm CPD}_{i,j}(\lambda+\delta \lambda).
\end{equation}
\noindent 
The CPD channel map was then co-added with the partially extincted CSD channel map. The data products that contribute to this overall process are illustrated as Case 1 in Fig. \ref{fig:cube-combiner}. In the event that the CPD phase angle $\theta$ was $>90$°, the roles are reversed and the CPD was partially extincted by foreground dust and gas in the CSD cavity wall (Case 2 in Fig. \ref{fig:cube-combiner}) .  In this case the $\tau$ map of the CSD was used to extinct the CPD.

\subsection{SimMETIS}

To simulate the capabilities of the METIS instrument we used the SimMetis software package \footnote{https://github.com/astronomyk/SimMETIS release v0.2 (retrieved 19 Feb 2019)} based on SimCADO.  The telescope altitude is set to 3060 m at latitude -24.59° and longitude -70.19°. The telescope temperature is 282.15 K. The detector pixel scale is 8.2 mas\footnote{In reality the LMS mode FOV is  0.58 × 0.93'' cut into 28 slices of 0.021 x 0.93''. The slices are projected onto a detector with plate scale 0.0082'' per pixel. Hence as the PSF is undersampled in the across-slice direction an observation will require a series of exposures with dithers/offsets or rotations, from which a spatially fully sampled data cube will be reconstructed.  The SimMetis pixel scale thus represents an ideal data reduction scenario.}, corresponding to $\sim1.1$ au at 108 pc  \citep{Brown2021}.  The exposure time and number of exposures has been varied.  We consider a detector integration time (DIT) of 10 s and a number of DITs (NDIT) for total exposure times totalling 10s, 60s, 1h, and  4h.  A longer DIT is not adopted to prevent saturation of the detector.   From the maximum possible elevation of HD 100546 and altitude of the ELT we adopt for observing parameters an airmass of 1.5. The atmospheric conditions are "median".

\section{Results} \label{sec:results}

\begin{figure*}
  \centering
  \begin{subfigure}{\textwidth}
    \centering
    \includegraphics[width=\textwidth]{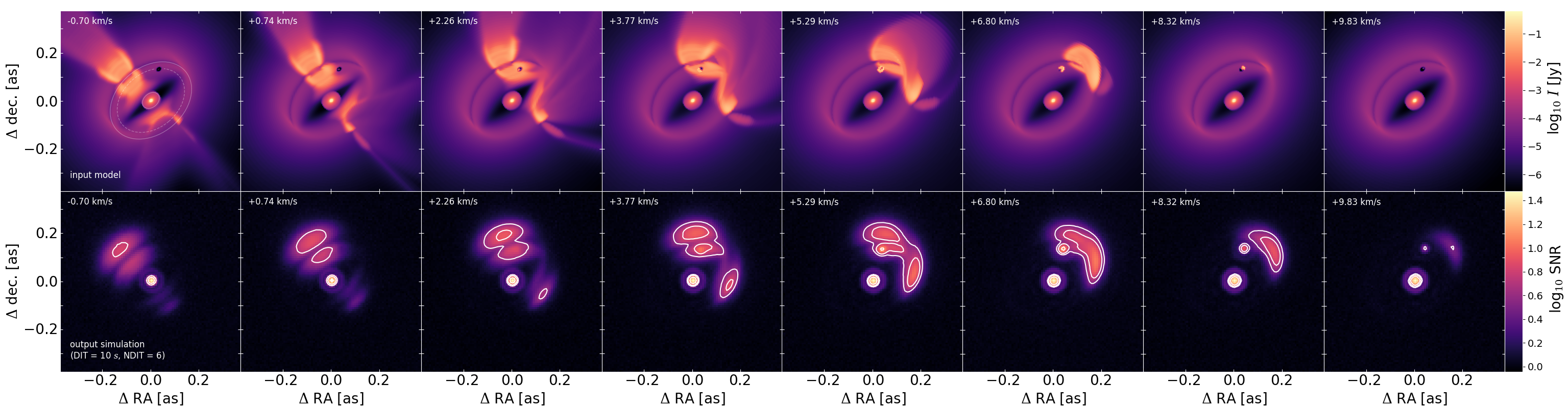}%
    \caption{$\chi_{\rm RT} = 2\times10^6$, $\theta = 45$°, $R_{\rm out} = $ R$_{\rm Hill}$}
    \label{fig:1234_PA45_LOWCHI5a}
  \end{subfigure} 
  \vskip\baselineskip
  \vspace{-2ex}
  \begin{subfigure}{\textwidth}
    \centering
    \includegraphics[width=\textwidth]{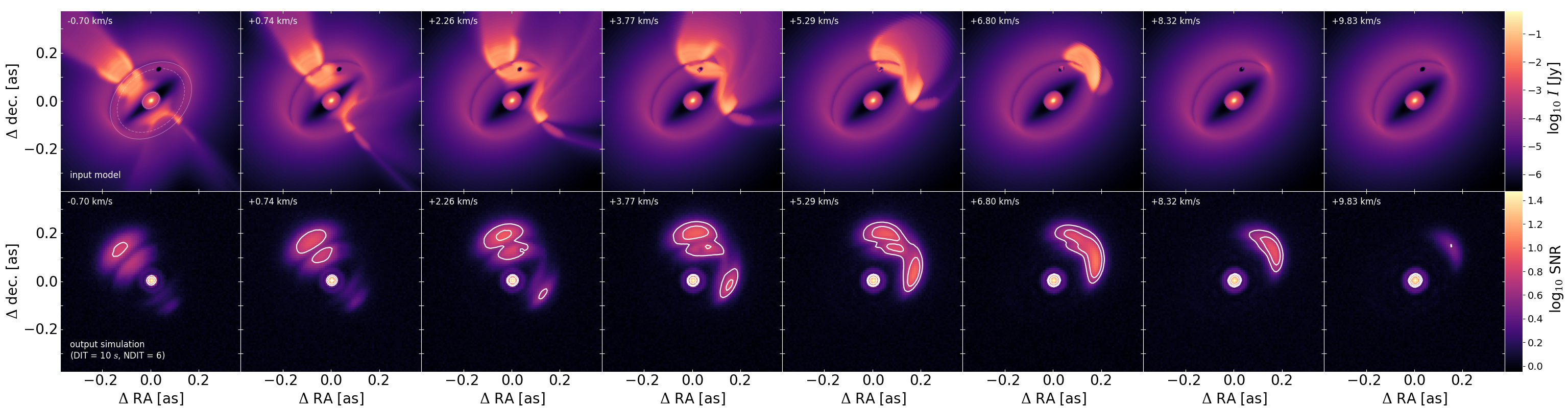}%
    \caption{$\chi_{\rm RT} = 10^5$, $\theta = 45$°, $R_{\rm out} = $ R$_{\rm Hill}$}
    \label{fig:1234_PA45_LOWCHI5b}
  \end{subfigure} 
  \caption{Synthetic channel maps of the v(1-0)P08 line (4.73587 \textmu m) of the HD 100546 CSD + CPD model for the high external irradiation case (a) and for the low external irradiation case (b).  In both cases the CPD outer radius $R_{\rm out} = $ R$_{\rm Hill}$ (1.62 au).  The \textsc{ProDiMo} model used as input for SimMetis is in the top row of each subfigure.  The ellipses in the upper-leftmost panel indicate the inner and outer edges of the gap at 4 and 19 au, and the companion orbit at 15 au (dotted line).  The corresponding simulated METIS observation panels represent 6 detector integrations of 10 s each in the bottom row of every subfigure. The white contour lines represent signal-to-noise ratios of 3, 5, and 10. The velocity offset is relative to the stellar reference frame.}
\label{fig:1234_PA45_LOWCHI5}
\end{figure*}

\begin{figure*}
  \includegraphics[width=\textwidth]{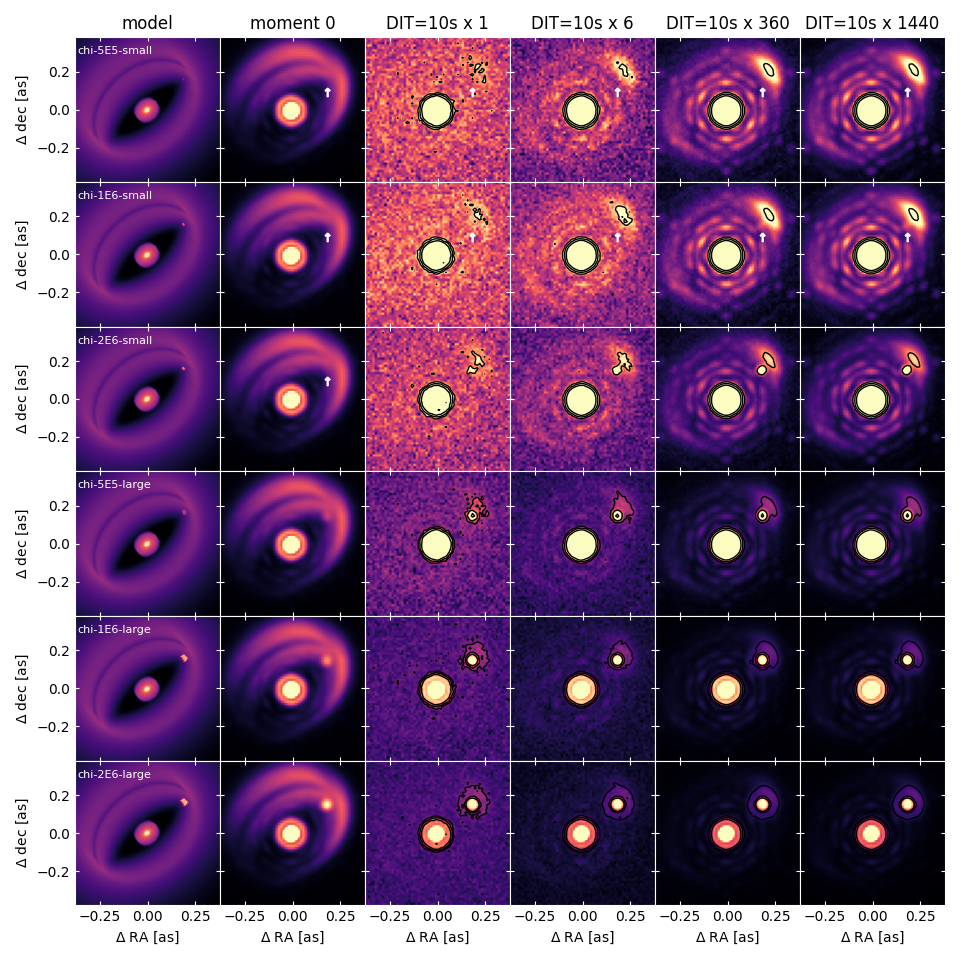}
      \caption{Synthetic channel maps of v(1-0)P08 line  (4.73587 \textmu m) from the CPDs of varying physical extent and background irradiation. The CPD is placed at maximum elongation ($\theta = 90$°) The leftmost column contains the input model at the extracted velocity (+6.8 km s$^{-1}$ relative to the stellar reference frame). The asymmetry in the moment 0 map in column 2 is due to telluric contamination.  The third to sixth columns include the synthetic channel maps for an increasing series of total detector integration times indicated at the top of each column.  The black contours indicate a count value excess relative to the background of $25\%$, $50\%$, and $100\%$. The color range of each image has been logarithmically scaled with boundaries corresponding to the minimum and maximum values found outside of the central half of the stellar PSF. Where it is not apparent by eye the position of the CPD is indicated with a white vertical arrow. The six-fold symmetry apparent in longer integration times is a result of the ELT point spread function acting on the central star. The colormap normalization has been configured to maximize the visibility of the CPD.}
      \label{fig:big-grid}
\end{figure*}

 \begin{figure*}
        \centering
        \begin{subfigure}[b]{0.49\textwidth}
            \centering
            \includegraphics[width=\textwidth]{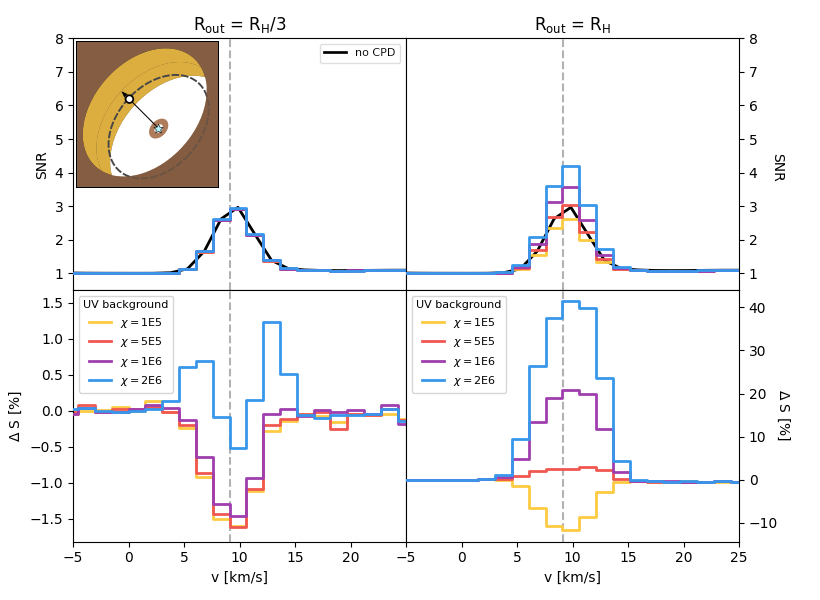}
            \caption[Network2]%
            {{\small Position Angle + 0°}}    
            \label{fig:subsnr0}
        \end{subfigure}
        \hfill
        \begin{subfigure}[b]{0.49\textwidth}  
            \centering 
            \includegraphics[width=\textwidth]{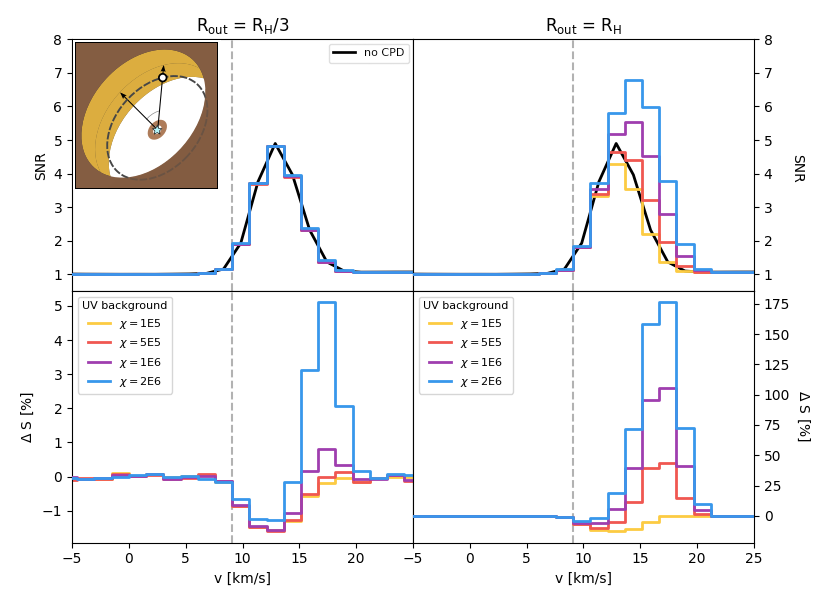}
            \caption[]%
            {{\small Position Angle + 45°}}    
            \label{fig:subsnr1}
        \end{subfigure}
        \vskip\baselineskip
        \begin{subfigure}[b]{0.49\textwidth}   
            \centering 
            \includegraphics[width=\textwidth]{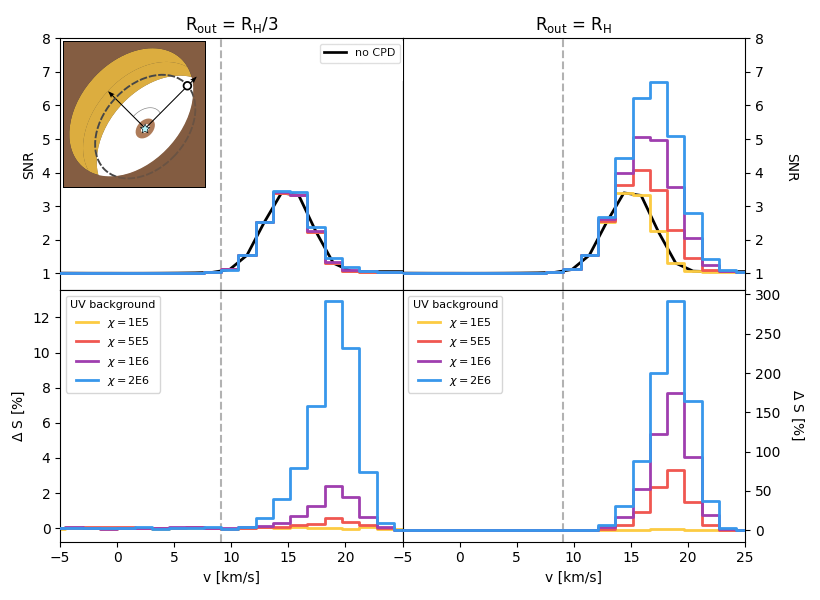}
            \caption[]%
            {{\small Position Angle + 90°}}    
            \label{fig:subsnr2}
        \end{subfigure}
        \hfill
        \begin{subfigure}[b]{0.49\textwidth}   
            \centering 
            \includegraphics[width=\textwidth]{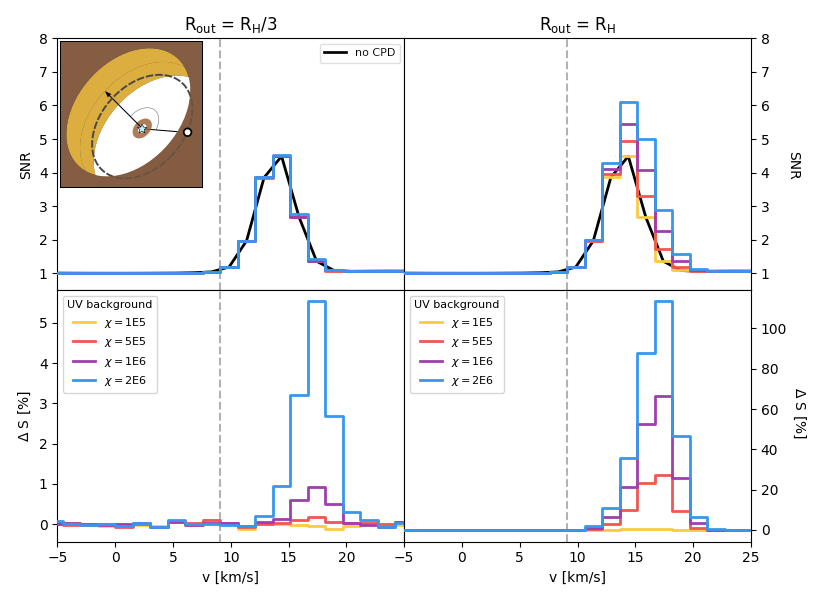}
            \caption[]%
            {{\small Position Angle + 135°}}    
            \label{fig:subsnr3}
        \end{subfigure}
        \caption[ test ]
        {\small $^{12}$CO line profiles represented as the signal-to-noise ratio (SNR) of the aperture centered on the CPD (\textit{top row of each subfigure}) compared to the case of no CPD (black line) for the CPDs with varying levels of background UV irradiation $\chi$.  Excess or deficit in flux $\Delta S$ in the aperture relative to a no-CPD model (\textit{bottom row of each subfigure}). Results are shown for a companion position angle relative to the disk semiminor axes of (a) 0$^{\circ}$, (b) 45$^{\circ}$, (c) 90$^{\circ}$, and (d) 135$^{\circ}$.  The signal present in the case of no CPD is indicated by the black lines, and arises purely from circumstellar disk emission.  The rest frame of the star is indicated by the vertical dashed gray lines. A negative $\Delta S$ indicates that the CPD is seen in absorption. Note that the vertical axis scaling can differ significantly between the left and right sides of each subfigure.} 
        \label{fig:SNR001}
    \end{figure*}

At the CO line centers we find the CPD flux is 10$^2$-10$^4$ greater than the planet+CPD continuum emission at \mbox{4.6-4.9 \textmu m} depending on the magnitude of external irradiation.  Over this wavelength range the CPD dust contributes at most 10$\%$ of the combined planet+CPD emission. For the reference case we find the peak CO line flux near \mbox{4.83 \textmu m} is of the same order as the entire CSD disk-integrated continuum flux. This is evident in Fig. \ref{fig:sed-with-lines}.

The flux of the CPD is extracted by placing a circular aperture at the position of its centroid and summing across all pixels in the aperture. The aperture has a radius of 3 pixels and is sized to include $>99.5\%$ of the emission originating from the CPD.   Expressing the statistical significance of the CPD emission is complicated by contamination (blending) of CSD emission within the aperture.  Despite the Doppler shifting of the CPD line emission we find that some contamination from circumstellar disk emission occurs in every case for a CPD radial separation of 15 au.  Fluxes are extracted from the spectral channel with the maximum contrast with the CSD.  A value of 100$\%$ would indicate a doubling of the emission within the aperture relative to a no-CPD model. With real data a similar analysis would require the subtraction of modeled CSD emission.  

A lower dust-to-gas ratio allows UV radiation to penetrate more deeply into the CPD and heat the gas, increasing the CO line emission by up to a factor $\sim2.5$.  A CPD for a planet mass as low as \mbox{1 M$_{\rm J}$} is still present as a localized excess emission of 10-60$\%$.  If either the CPD gas mass is reduced by a factor 1000 (to \mbox{10$^{-4}$ M$_{\rm J}$)}, the UV background is reduced by a factor 20, or the CPD gas component radius is reduced to R$_{\rm H}/3$, the signal of the CPD is not present as an excess of more than 10$\%$ and becomes more challenging to detect.  A phase angle of 90° (maximum elongation) increases the contrast with circumstellar disk emission by a factor 2 relative to \mbox{$\theta = 45^{\circ}$}.  Conversely the CPD is not detectable at \mbox{$\theta = 180^{\circ}$}, where the CPD is spatially coincident with the optically thick line-emitting region of the CSD (\mbox{$\tau > 1300$}).

The CPD $^{12}$CO line strength is found to be coupled only weakly to the planetary luminosity.  Although the inner rim of the CPD is heated by the planet, the corresponding effective emitting area is negligible. The majority of the line emission originates instead from the disk surface. Background UV irradiation of the CPD efficiently increases gas temperatures in a thin surface layer up to \mbox{4000 K}. This can be seen in Fig. \ref{fig:cpd-tgas}. The resulting emission region is comparable to the total disk surface area. The main emission region is highlighted in Fig. \ref{fig:cpd-line-origin}.  The line luminosity is thus closely proportional to the physical extent of the circumplanetary gas and the intensity of the UV background. In the ``best case" scenario for detection, a Hill sphere-filling CPD with background \mbox{$\chi = 2\times10^6$} produces spatially resolved emission with peak signal-to-noise ratio \mbox{SNR = 5} in 60 s of total exposure. If the external irradiation is reduced by a factor 20 the CPD becomes difficult to distinguish from background circumstellar disk emission. Likewise if the CPD outer radius is truncated to R$_{\rm Hill}/3$, detection of the CPD gas requires the most extreme case of background irradiation to ensure the CPD emission can be distinguished from the background CSD gap wall at a level greater than 10$\%$ localized excess. 

We have produced a series of synthetic channel maps for each combined CSD+CPD model representing plausible observational parameters.  Synthetic channel maps of the combined CSD+CPD for the high- and low-levels of background radiation can be found in Fig. \ref{fig:1234_PA45_LOWCHI5a} and \ref{fig:1234_PA45_LOWCHI5b} respectively. The CPD is both spatially and spectrally resolved in Fig. \ref{fig:1234_PA45_LOWCHI5a} owing to the line-of-sight velocity offset with the gap wall emission surface. Interestingly the presence of the CPD can be inferred in Fig. \ref{fig:1234_PA45_LOWCHI5b} even in the absence of detectable line emission. The CPD dust continuum absorption produces a silhouette on the background of the CSD gap wall line emitting region if the external irradiation is sufficiently low. This effect is clearly demonstrated in Fig. \ref{fig:1234_PA45_LOWCHI5b} at \mbox{+2.26 km s$^{-1}$} and  \mbox{+3.77 km s$^{-1}$} where the position of the CPD is apparent in absorption. 

The greatest contrast between the CPD and CSD is found at maximum elongation ($\theta = 90^{\circ}$) when the relative Doppler shifting between the planet and star are maximal. Synthetic channel maps for the high- and low-levels of background radiation for both small \mbox{($R_{\rm out} = $ R$_{\rm H}/3$)} and large \mbox{($R_{\rm out} = $ R$_{\rm H}$)} CPD models at maximum elongation can be found in Fig. \ref{fig:big-grid}. In all cases where $R_{\rm out} = $ R$_{\rm H}$ the CPD emission can be easily distinguished from the circumstellar disk by eye. 

\begin{figure}
  \includegraphics[width=0.5\textwidth]{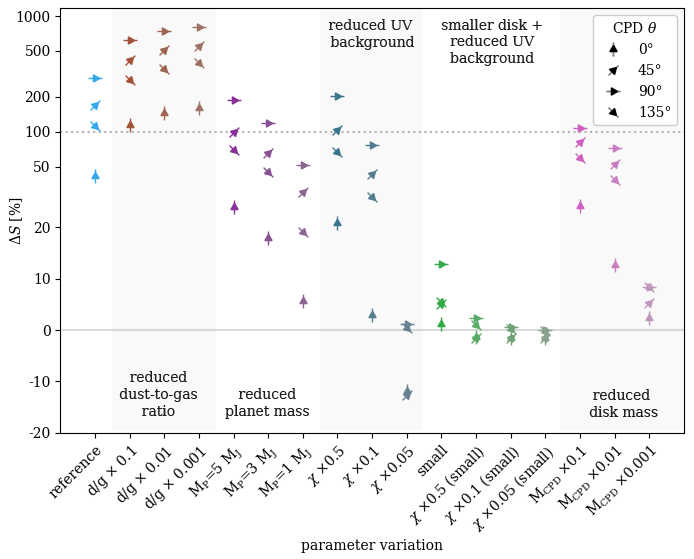}
      \caption{CPD induced excess/deficit flux in aperture for all CPD model configurations and phase angles $\theta$. Fluxes are extracted from the spectral channel with the maximum contrast with circumstellar disk emission. A value below 0 indicates the CPD is seen in absorption.  A value larger than 100 indicates that the aperture flux is more than double that of a no-CPD model. The variation of parameters are relative to the reference CPD with planet mass \mbox{$M_{\rm P} = 10$M$_{\rm J}$}, \mbox{M$_{\rm CPD} = 0.01 M_{\rm P}$}, \mbox{$\chi_{\rm RT}$ = $2\times10^6$}, \mbox{$d/g$ = $10^{-3}$}.  A label "small" indicates that the CPD outer radius has been truncated to one third of the planet Hill radius. The angle of the triangle markers indicates the phase angle of the CPD relative to the circumstellar disk northern minor axes.}
      \label{fig:summary}
\end{figure}

The detectability of the CPD represented both in terms of the aperture SNR and as an excess or deficit in emission relative to a no-CPD model for different phase angles can be found in in Fig. \ref{fig:SNR001}. The CPD continuum absorption silhouette effect is also clear in Fig. \ref{fig:SNR001} (a) and (b) for $\theta = 0^{\circ}$ and $45^{\circ}$, respectively, with a localized reduction in flux relative to a smooth axisymmetric of $\sim10\%$. We summarize our parameter exploration in terms of the significance of the CPD emission as an excess or deficit in the flux relative to a no-CPD model in Fig. \ref{fig:summary} for all model configurations.

Given the instantaneous spectral coverage of the METIS instrument as many as eight $^{12}$CO v=1-0 lines may be present in a single observation. Depending on the magnitude of telluric contamination the eventual signal-to-noise of a CPD detection may thus be increased by a factor $\sim2.5-3$ over what we demonstrate in e.g. Fig. \ref{fig:SNR001}.

As the C/O ratio of material accreting from the outer circumstellar disk onto the CPD may differ substantially from abundances corresponding to the ISM (e.g. \citet{Ansdell2016,Zhang2019CO}), we considered also the case of a carbon depletion by a factor 10.  Given that the $^{12}$CO v=1-0 lines are highly optically thick, we find that a 10$\times$ depletion of C results in peak line fluxes being reduced by $\sim20\%$ relative to the value corresponding to standard ISM abundances.

\section{Discussion} \label{sec:discussion}

We have found that a sufficiently large and externally irradiated CPD can produce a signal which could be detected even when the entire gap region of the CSD would be spatially unresolved. Although without a priori knowledge of the line-of-sight velocity offset of the CPD signal w.r.t. the circumstellar disk emission, multi-epoch observations would be required to deduce Keplerian motion of the signal originating from the CPD. In the most extreme case we find the CPD peak line luminosity approaches \mbox{$\sim40\%$} of the disk-integrated circumstellar emission at the same wavelength (see Fig. \ref{fig:sed-with-lines}). This is comparable with the magnitude of the candidate CPD signal observed in $^{12}$CO P26 line emission discussed in \citet{Brittain2013II}.  The  centrally peaked CO line morphology arises as a result of the emission originating from across the entire disk surface at relatively low Keplerian velocities.

Dust depletion of CPDs due to grain drift is predicted to occur on relatively short timescales \citep{Zhu2018,Rab2019}, frustrating attempts to detect CPDs by continuum emission. This is consistent with the non-detection of CPD-like dust emission in several pre-transitional disk cavities. However we find that a CPD with severely depleted dust ($d/g\sim10^{-6}$) can still be readily detected in CO emission, while being simultaneously impossible to detect in continuum emission with e.g. ALMA.  Other pre-transitional disks around UV-bright Herbig Ae/Be stars  e.g. UX Tau A , HD34282, HD97048 ,CQ Tau, MWC 758 \citep{Andrews2011,Plas2017,Plas2017b,Gabellini2019,Calcino2020} offer similar opportunities to search for CO emission from externally irradiated CPDs.

The possibility to detect the CPD in absorption against the CSD gap wall represents an unique observational scenario allowing the dust properties of the CPD to be revealed.    In this sense the CPD is comparable to an externally irradiated proplyd such as Orion 121-1925 \citep{McCaughrean1996}.  In the event that the magnitude of the external radiation is as low or lower than the minimum we have considered, this possibility will allow for the physical extent, dust properties, and minimum dust mass of the CPD to be estimated. However, even in the most optimistic case the spatially localized reduction in flux relative to a no-CPD model ($\sim10\%$) is minor, and spatial inhomogeneity of background circumstellar disk CO emission could easily obscure such a signal.  Thus in practice it may be very difficult to achieve lest the position of the planet and CPD were known a priori.

\section{Conclusions} \label{sec:conclusions}

We have used thermochemical disk models to produce synthetic channel maps that represent $^{12}$CO line observations of the HD 100546 system in an effort to determine the plausibility of detecting CPDs in transition disk gaps with METIS.  We list our key findings:

   \begin{enumerate}
   
      \item  Fundamental $^{12}$CO ro-vibrational line emission from the gas component of a circumplanetary disk can potentially be detected in only 60 s of detector integration time with ELT/METIS
      \item Visibility of the CPD depends strongly on the level of external irradiation and the physical extent of the disk, favoring massive ($\sim 10$ M$_{\rm J}$) planets and spatially extended disks with radii approaching the planetary Hill radius.  
      \item The majority of $^{12}$CO line emission originates from across the entire disk surface, and thus the CO line profiles are centrally peaked. The planetary luminosity does not play a significant role in exciting the $^{12}$CO line emission.
      \item Massive, UV-bright star systems with pre-transitional disks are ideal candidates to search for CO-emitting CPDs with ELT/METIS.
   \end{enumerate}

\noindent
The capabilities of METIS represent a critical component of the multi-pronged effort to unveil the processes of giant planet and moon formation.  If CPDs prove to be strongly dust-depleted, gas line observations will play a critical role in this endeavour. 

\begin{acknowledgements}
      The research of N.O. and I.K. is supported by grants from the Netherlands Organization for Scientific Research (NWO, grant number 614.001.552) and the Netherlands Research School for Astronomy (NOVA). CHR acknowledges the support of the Deutsche Forschungsgemeinschaft (DFG, German Research Foundation) Research Unit ``Transition discs'' - 325594231. Ch.R is grateful for support from the Max Planck Society. N.O. would like to thank B.R. Brandl and R. van Boekel for helpful discussions.  This research has made use of NASA's Astrophysics Data System Bibliographic Services.  This research made use of Astropy 3 a community-developed core Python package for Astronomy \citep{astropy:2013,astropy:2018})  This research has also used Numpy \citep{numpy},  Matplotlib \citep{matplotlib}, Scipy \citep{scipy}, and Prodimopy \url{https://gitlab.astro.rug.nl/prodimo/prodimopy}. . 
\end{acknowledgements}

%
%
\bibliographystyle{aa} 
\bibliography{refs.bib} 

\begin{appendix}


\section{Properties of the reference CPD model}

\begin{figure*}
\centering

\begin{subfigure}[b]{0.49\textwidth}
\hspace*{-0.3cm}   
  \includegraphics[width=1\linewidth]{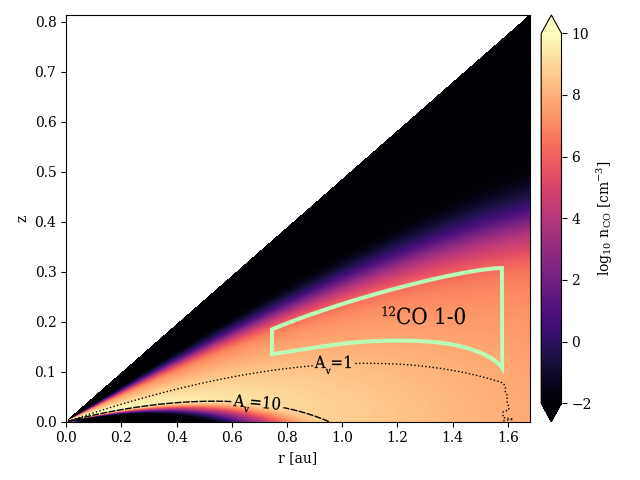}
      \caption{CPD CO abundance and line emitting region.}
      \label{fig:cpd-line-origin}
\end{subfigure}
\hfill
\begin{subfigure}[b]{0.49\textwidth}
  \includegraphics[width=1\linewidth]{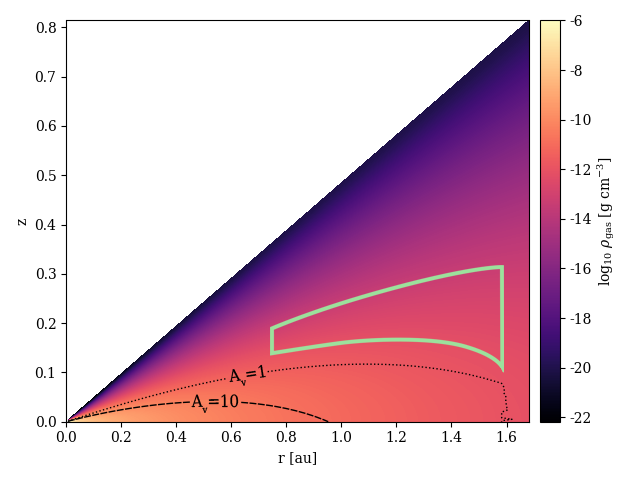}
      \caption{CPD gas density.}
      \label{fig:cpd-rhog}
\end{subfigure} 
\vskip\baselineskip
\begin{subfigure}[b]{0.49\textwidth}
  \includegraphics[width=1\linewidth]{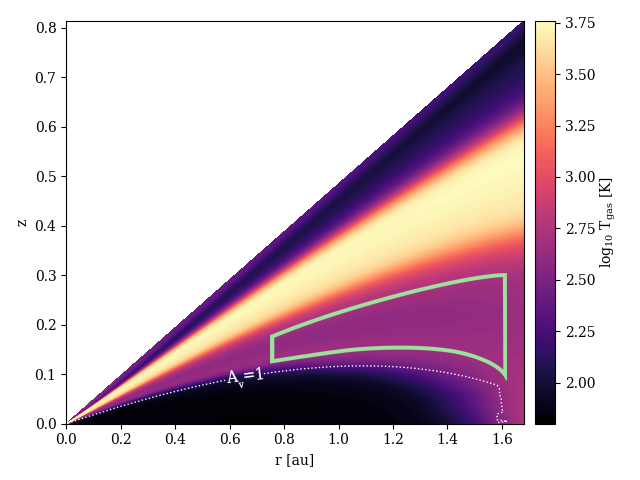}
  \caption{CPD gas temperature.}
      \label{fig:cpd-tgas}
\end{subfigure}
\hfill
\begin{subfigure}[b]{0.49\textwidth}
  \includegraphics[width=1\linewidth]{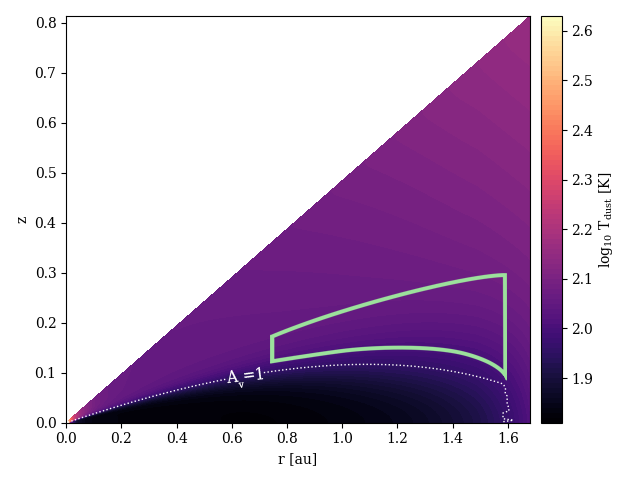}
  \caption{CPD dust temperature.}
      \label{fig:cpd-tdust}
\end{subfigure}

\vskip\baselineskip
\begin{subfigure}[b]{0.49\textwidth}
  \includegraphics[width=1\linewidth]{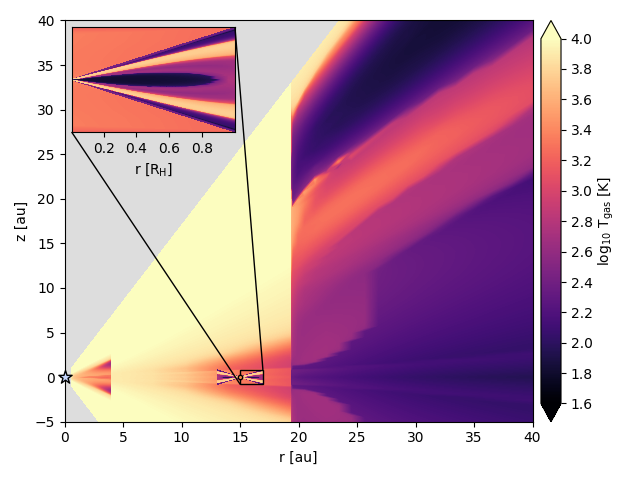}
  \caption{CPD and CSD gas temperature.}
      \label{fig:combo-tgas}
\end{subfigure}
\hfill
\begin{subfigure}[b]{0.49\textwidth}
  \includegraphics[width=1\linewidth]{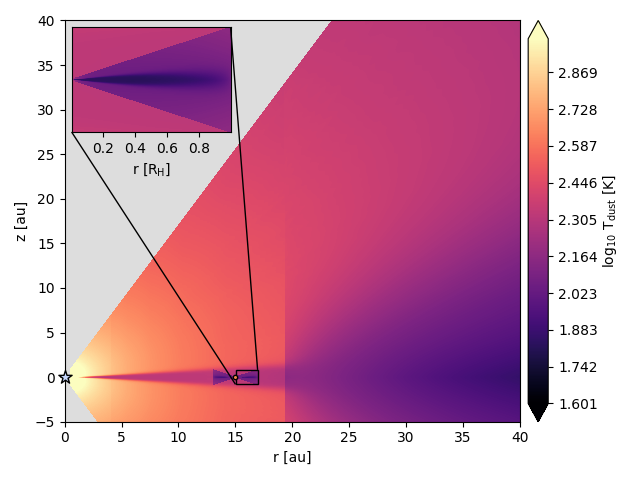}
  \caption{CPD and CSD dust temperature.}
      \label{fig:combo-tdust}
\end{subfigure}

\caption{Properties of the reference CPD with $\chi = 2\times10^6$ and $R_{\rm out} = R_{\rm Hill}$ (a,b,c,d). The majority of the $^{12}$CO v=1-0 line emission originates from within the green ``box" region in panels a-d.  The dotted contour lines in panels (a,b,c,d) represent minimum optical extinction A$_{\rm V}$ surfaces of 1 (dotted line), 10 (dashed line) in the radial or vertical direction. The gas and dust temperature structure of the CPD is contrasted for illustrative purposes with that of the circumstellar disk in panels (e) and (f).}

\end{figure*}

The $^{12}$CO abundance and v=1-0 line emitting region of the reference CPD can be found in Fig. \ref{fig:cpd-line-origin}.  The gas temperature and density structure can be found in Fig. \ref{fig:cpd-tgas} and  \ref{fig:cpd-rhog}.  Small dust grains in the upper layers of the CPD absorb external UV photons but cool efficiently to radiative equilibrium.  This heats the surrounding gas which in turn cools predominantly by the relatively inefficient H$_2$O and CO rotational and vibrational emission. Hence the gas and dust temperature are not closely coupled in this hot surface layer. The parameters of the CPD models which are common to all models described in Table \ref{tab:cpd-properties} are listed in Table \ref{tab:cpds}.

\section{CPD line emission strength relative to the continuum}

\begin{figure}[H]
  \includegraphics[width=1\linewidth]{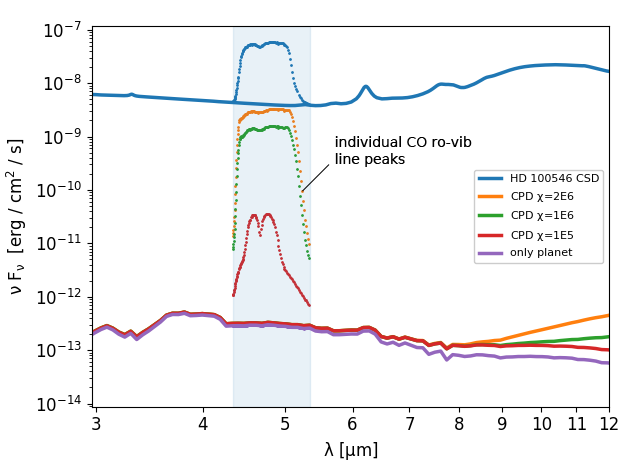}
      \caption{Comparison of the estimated CPD and CSD $^{12}$CO line fluxes (dots) for varying intensities of background UV radiation with the continuum SED of the CSD and CPDs (solid lines). Only $^{12}$CO lines are plotted.}
      \label{fig:sed-with-lines}
\end{figure}

The estimated peak emission strength of each of the modeled $^{12}$CO lines is shown in Fig. \ref{fig:sed-with-lines} for the circumstellar disk, three cases of varying background radiation incident on the CPD, and the case of a planet without CPD.

\section{Spatially unresolved CPD emission}

\begin{figure}[H]
  \includegraphics[width=1\linewidth]{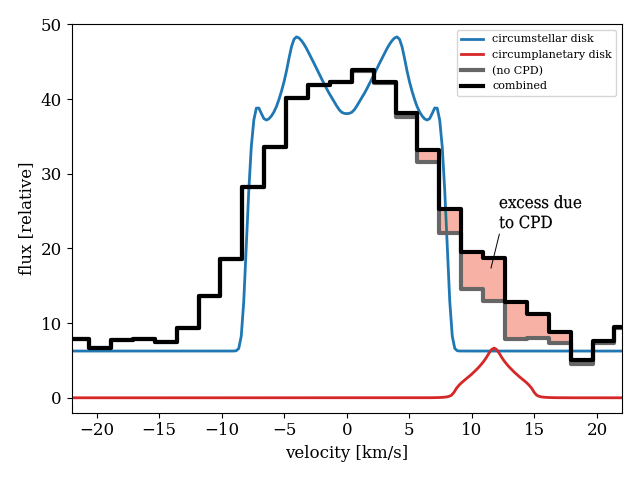}
      \caption{Synthetic $^{12}$CO v(1-0)P26 (4.9204 \textmu m) line profiles of the circumstellar disk (blue) and circumplanetary disk (red) from the \textsc{ProDiMo} radiative transfer.  The combined and degraded signal corresponds to the spectral resolution of the CRIRES observations performed by \citet{Brittain2014, Brittain2019} during multi-epoch monitoring of the HD 100546 system. The light shaded red region indicates the contribution of the CPD.}
      \label{fig:unresolved}
\end{figure}

We find that the CPD CO line emission can in certain cases be sufficient to be detectable even when spatially unresolved.  In Fig. \ref{fig:unresolved} we demonstrate a synthetic P26 line profile in the event that the disk would be spatially unresolved.  The spectral resolution and noise level of the combined CSD (blue line) + CPD (red line) line profile has been degraded to correspond to the excess CO emission observed with CRIRES and discussed in \citet{Brittain2014, Brittain2019} (black line).  The corresponding line profile in the absence of a CPD is indicated by the gray line.  While the CPD-induced excess is notable, without a priori knowledge of the properties and orbital phase of the CPD, spectral decomposition would be implausible without long-term monitoring of the signal.

\section{How much stellar radiation reaches the CPD?} \label{sec:appendix:gapUV}

\begin{figure}[H]
  \includegraphics[width=\textwidth/2]{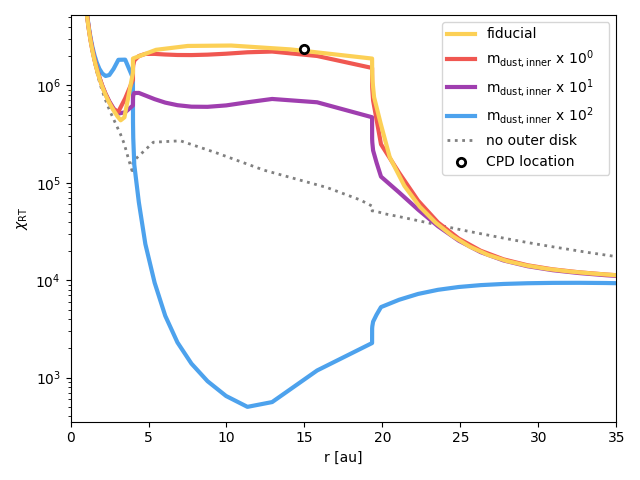}
      \caption{Midplane $\chi_{\rm RT}$ extracted from the 2D radiative transfer of the ProDiMo HD 100546 disk model for the reference case, and in the case of the dust mass in the gap being equal to 1, 10, and 100$\times$ the amount of dust in the inner disk.  The case of no outer disk is included as the gray dashed line to highlight the role of back-scattering off the gap outer wall.}
      \label{fig:midplane-chi}
\end{figure}

In our reference model we have assumed the gap to be effectively empty with a correspondingly negligible optical depth.  As a verification step we perform a series of radiative transfer simulations in which we populate the gap with additional dust and gas to explore the potential for extinction of stellar UV towards the CPD and of CPD emission towards the observer. Several cases of increased gap dust mass are displayed in Fig. \ref{fig:midplane-chi}. Non-neglible extinction of the stellar UV begins to occur if the gap is populated by 10$\times$ the dust mass inferred to be present in the inner disk. With a gap dust mass of $10 \times m_{\rm dust, inner}$ the midplane FUV intensity $\chi_{\rm RT}$ falls to $30\%$ of the reference value at the position of the CPD.  The SED of the CSD in this case is still consistent with observations, increasing the flux at 4 \textmu m by no more than $\sim 4\%$.  If the gap dust mass is increased to 100$\times$ the inner disk dust mass, the SED begins to diverge significantly from WISE, ISO-SWS, and VISIR observations at 2-11 \textmu m and is ruled out by observations. Hence we consider it plausible that additional dust in the gap may reduce the CPD-incident FUV flux by at least $\sim70\%$ without becoming inconsistent with observations.

We found in Sect. \ref{sec:cpd-chi} that upwards of 95$\%$ of the UV radiation incident on the CPD originates from back-scattering off of dust in the gap outer wall. While radiative transfer in \textsc{ProDiMo} is calculated only using isotropic scattering of photons, at wavelengths $\sim 0.1$ \textmu m molecular gas can act as an effective Rayleigh-scatterer. Furthermore, stellar Ly-$\alpha$ photons may be resonantly scattered downwards into a gap by atomic hydrogen in the upper layers of the circumstellar disk, although this phenomenon is also not included in the 2D radiative transfer \citep{Bethell2011}. This leads us to believe that isotropic incidence of external UV is a reasonable approximation in this case. Nevertheless we consider the case where the external radiation has been decreased by a factor 20 under the assumption that scattering of UV photons would be particularly inefficient or that significant quantities of undetected dust fills the gap.


\end{appendix}

\end{document}